\documentclass[12pt]{article}
\usepackage{epsf,amsfonts,amssymb,epsfig,amsmath}
\renewcommand{\text}[1]{#1}

\newcommand{\be}{\begin{equation}}
\newcommand{\ee}{\end{equation}}
\newcommand{\ben}{\begin{displaymath}}
\newcommand{\een}{\end{displaymath}}
\newcommand{\bea}{\begin{eqnarray}}
\newcommand{\eea}{\end{eqnarray}}
\newcommand{\bean}{\begin{eqnarray*}}
\newcommand{\eean}{\end{eqnarray*}}
\newcommand{\nn}{\nonumber \\}
\newcommand{\ba}{\begin{array}}
\newcommand{\ea}{\end{array}}
\newcommand{\bi}{\begin{itemize}}
\newcommand{\ei}{\end{itemize}}

\newcommand{\reef}[1]{(\ref{#1})}

\newcommand{\bbR}{{\mathbb{R}}}

\begin{document}

\makeatletter
\renewcommand{\theequation}{\thesection.\arabic{equation}}
\@addtoreset{equation}{section}
\makeatother

\baselineskip 18pt

\begin{titlepage}

\vfill

\begin{flushright}
Imperial/TP/2009/JG/03\\
DESY 09-067\\
\end{flushright}

\vfill

\begin{center}
   \baselineskip=16pt
   {\Large\bf  Solutions of type IIB and D=11 supergravity\\with Schr\"odinger$\left( z\right)$ symmetry}
  \vskip 1.5cm
      Aristomenis Donos$^1$ and Jerome P. Gauntlett$^2$\\
   \vskip .6cm
      \begin{small}
      $^1$\textit{DESY Theory Group, DESY Hamburg\\
        Notkestrasse 85, D 22603 Hamburg, Germany}
        \end{small}\\*[.6cm]
        \begin{small}
      $^2$\textit{Theoretical Physics Group, Blackett Laboratory, \\
        Imperial College, London SW7 2AZ, U.K.}
        \end{small}\\*[.6cm]
      \begin{small}
      $^2$\textit{The Institute for Mathematical Sciences, \\
        Imperial College, London SW7 2PE, U.K.}
        \end{small}\\*[.6cm]
   \end{center}

\vfill

\begin{center}
\textbf{Abstract}
\end{center}

\begin{quote}
We construct families of supersymmetric solutions of type 
IIB and $D=11$ supergravity that
are invariant under the non-relativistic Schr\"odinger$\left( z\right)$ 
algebra for various values of the dynamical
exponent $z$. The new solutions are based on five- and 
seven-dimensional Sasaki-Einstein manifolds, respectively, 
and include supersymmetric solutions with $z=2$.
\end{quote}

\vfill

\end{titlepage}
\setcounter{equation}{0}


\section{Introduction}
An interesting development in string/M-theory is the possibility of using holographic ideas
to study condensed matter systems. Starting with \cite{Son:2008ye,Balasubramanian:2008dm},
one focus has been on non-relativistic systems with
Schr\"odinger symmetry, a non-relativistic version of conformal symmetry.
The corresponding Schr\"odinger algebra is generated by Galilean transformations, an
anisotropic scaling of space (${\bf x}$) and time ($x^+$) coordinates given
by ${\bf x}\to \mu {\bf x}$, $x^+\to \mu^2 x^+$, and an additional special conformal transformation.
More generally, one can consider systems invariant under what we shall call Schr\"odinger$\left( z\right)$
(or Sch$\left( z\right)$) symmetry, where one maintains the Galilean transformations,
but allows for other scalings, ${\bf x}\to \mu {\bf x}$, $x^+\to \mu^z x^+$, with
$z$ the ``dynamical exponent'', and, in general, sacrifices the special conformal transformations.
In this notation the Schr\"odinger algebra is Sch$(2)$.
The full set of commutation relations for $Sch\left( z\right)$ are written down in e.g. \cite{Balasubramanian:2008dm}.

Various solutions of type IIB supergravity and $D=11$ supergravity have been constructed
that are invariant under Sch$\left( z\right)$ symmetry, for different values of $z$.
The type IIB solutions of \cite{Herzog:2008wg}-\cite{Donos:2009en}
can be viewed as deformations of the supersymmetric $AdS_5\times SE_5$ solutions, where $SE_{5}$ is a five-dimensional Sasaki-Einstein space,
and should be
holographically dual to non-relativistic systems with two spatial dimensions. Similarly, there
are deformations of the $AdS_4\times SE_7$ solutions\footnote{Note that deformations of $AdS_5$ solutions
of $D=11$ supergravity were studied in \cite{Colgain:2009wm}.} of $D=11$ supergravity,
where $SE_{7}$ is a seven-dimensional Sasaki-Einstein space,
that are invariant under Sch$\left( z\right)$ and these should be dual to non-relativistic systems with a single spatial dimension
\cite{gkvw,Donos:2009en}.

The type IIB solutions constructed in \cite{Herzog:2008wg,Maldacena:2008wh,Adams:2008wt}
with $z=2$, and hence invariant under the larger Schr\"odinger algebra, are
based on a deformation in the three-form flux and do not preserve any supersymmetry \cite{Maldacena:2008wh}.
In \cite{Hartnoll:2008rs}
supersymmetric solutions of type IIB with various values of $z$ were constructed which are based on a metric deformation
and include supersymmetric solutions with $z=2$. However, it was argued that these supersymmetric solutions are unstable.
On the other hand it was shown that the instability can be removed by also switching on the three-form flux deformation, which then breaks supersymmetry.
In a more recent development a rich class of supersymmetric solutions of both type IIB and $D=11$ supergravity
were constructed in \cite{Donos:2009en} which have various values of $z\ge 4$ and $z\ge 3$, respectively (particular
examples of the $z=4$ and $z=3$ solutions were first constructed in \cite{Maldacena:2008wh} and \cite{gkvw}, respectively).

In this short note, we generalise the constructions in \cite{Donos:2009en}
for both type IIB and $D=11$ supergravity, finding new classes of
supersymmetric solutions with various values of $z$ including $z=2$.

{\bf Note Added:} In the process of writing this paper up, we became 
aware of \cite{Bobev:2009mw}, which also constructs 
some of the supersymmetric solutions of type IIB supergravity that we present 
in section 2.

\section{Solutions of type IIB supergravity}

Consider the general ansatz for the bosonic fields of type IIB supergravity given by
\bea
ds_{10}^{2}&= & \Phi^{-1/2}\left[2dx^+dx^-+h(dx^+)^2 +2Cdx^+ +dx_1^2+dx_2^2     \right] +\Phi^{1/2}ds^2(CY_3)\nn
F_{5}&= & dx^{+}\wedge dx^{-}\wedge dx_{1}\wedge dx_{2}\wedge d\Phi^{-1}+\,\ast_{CY_{3}}d{\Phi}\nn
 && -dx^{+}\wedge\left[\ast_{CY_{3}}dC+d\left({\Phi}^{-1}C\right)\wedge dx_{1}\wedge dx_{2}\right]\nn
G&= & dx^{+}\wedge W
\eea
where $G$ is the complex three-form and the axion and dilaton are set to zero. Here
$\Phi$, $h$ are functions, $C$ is a one-form and $W$ is a complex two-form all defined on the
Calabi-Yau three-fold, $CY_{3}$. Our conventions for type IIB supergravity \cite{Schwarz:1983qr,Howe:1983sra}
are as in \cite{Gauntlett:2005ww}.
One finds that all the equations of motion are satisfied provided that
\begin{align}
\nabla^{2}_{CY}\Phi&=0\nn
d\ast_{CY}dC & =0\nn
dW=d\ast_{CY}W & =0\nn
\nabla^{2}_{CY}h & =-|W|^2_{CY}\end{align}
where $|W|^2_{CY}\equiv (1/2!)W^{ij}W^\ast_{ij}$ with indices raised
with respect to the $CY$ metric.
Observe that when $C=h=W=0$ we have the standard D3-brane class
of solutions with a transverse $CY_3$ space.

If we choose the two-form $W$ to be primitive and have no $(0,2)$ pieces (i.e. just
$(2,0)$ and/or $(1,1)$ components),
on $CY_3$ then
the solutions generically preserve 2 supersymmetries\footnote{We note that one can
add a closed, primitive $(1,2)$-form $A$ on $CY_3$ to the three-form $G$ while still preserving the same amount of
supersymmetry. This changes two of the equations to
$\nabla^{2}_{CY}\Phi=-(1/2)|A|^2_{CY}$ and $d\ast_{CY}dC=i/2(W\wedge A^\ast-W^\ast\wedge A)$.
Such solutions will not, in general, admit a scaling symmetry, so we shall not consider them further here, however we note
that solutions with $W=0$ and Galilean symmetry were presented in \cite{Pal:2009np}.},
which is enhanced to 4 supersymmetries if the $CY_3$ is flat.
More specifically, we introduce the orthonormal frame
$e^+=\Phi^{-1/4}dx^+, e^-=\Phi^{-1/4}(dx^-+C+\frac{h}{2}dx^+), e^2=\Phi^{-1/4}dx^1,$ etc. and
choose positive orientation to be given by $e^{+-23}\wedge\rm{Vol}_{CY}$, where $\rm{Vol}_{CY}$ is the volume element on $CY_{3}$.
Consider first the special case that $C=h=W=0$. Then, as usual,
a generic $CY_3$ breaks 1/4 of the supersymmetry, while the harmonic function $\Phi$ leads
to a further breaking of 1/2, the Killing spinors satisfying the additional projection $\Gamma^{+-23}\epsilon=i\epsilon$.
Switching on $C,h,W$ we find that generically 
we need to also impose $\Gamma^+\epsilon=0$ and $\Gamma^{ij}W_{ij}\epsilon^c=0$.

We now specialise to the case that the $CY_{3}$ is a metric cone over a five-dimensional Sasaki-Einstein manifold $SE_5$,
$ds^2(CY_3)=dr^2+r^2ds^2(SE_5)$. In order to get solutions with Sch$(z)$ symmetry we now set
\bea
\Phi&=&r^{-4}\nn
C&=& r^{\lambda_{1}}\beta\nn
h&=& r^{\lambda_{2}}q\nn
W&=& d\left(r^{\lambda_{3}}\sigma\right)
\eea
where $q$ is a function, $\beta$ and $\sigma$ are, respectively, a real and a complex one-form
on $SE_5$, and $\lambda_i$ are constants which we will take to be positive.
The full solution now reads
\bea
ds_{10}^{2}&= & r^{2}\left[2dx^+dx^-+r^{\lambda_{2}}q\,(dx^+)^2 +2r^{\lambda_{1}}\,dx^+\beta +dx_1^2+dx_2^2     \right] +\frac{dr^{2}}{r^{2}}+ds^{2}\left(SE_{5} \right)\nn
F_{5}&= &4r^{3}\, dx^{+}\wedge dx^{-}\wedge dx_{1}\wedge dx_{2}\wedge dr+4\,\rm{Vol}_{SE_{5}}\nn
 && -dx^{+}\wedge\left[r^{\lambda_{1}+1}\,dr\wedge\ast_{SE_{5}}d\beta+\lambda_{1}r^{\lambda_{1}+2}\,\ast_{SE_{5}}\beta+d\left(r^{4+\lambda_{1}} \beta \right) \wedge dx_{1}\wedge dx_{2}\right]\nn
G&= & dx^{+}\wedge d\left(r^{\lambda_{3}}\sigma \right)~.
\eea
Generically, when $C,h,W\ne 0$, solutions with $\lambda_1+2=1+\lambda_2/2=\lambda_3\equiv z$ will be
Sch$(z)$ invariant. In particular, the scaling acts on the coordinates via
$(x^+,x^-,x_i,r)\to (\mu^zx^+,\mu^{2-z}x^-,\mu x_i,\mu^{-1}r)$
(for other transformations see \cite{Balasubramanian:2008dm}).
Observe that if we set $C=h=W=0$ then we have the standard $AdS_5\times SE_5$
solution of type IIB. Generically, when $C,h,W\ne 0$, we still need to impose the projections mentioned above in
order to preserve supersymmetry. Note in particular that, generically, half of the Poincar\'e supersymmetries
of the $AdS_5\times SE_5$ solution are preserved, while none of the special conformal supersymmetries are.
It would be interesting to explore 
special subclasses of solutions with enhanced supersymmetry, which occur,
for example, when the $CY_3$ is flat.

In \cite{Hartnoll:2008rs}, supersymmetric solutions with $W=C=0$, $h\ne0$ were constructed with
\begin{align}\label{qone}
\nabla_{SE}^{2}q+\lambda_2(4+\lambda_2)q=0
\end{align}
and give rise to solutions with $z=1+\lambda_2/2\ge 3/2$, with the bound only achievable for
$SE_5=S^5$. In particular supersymmetric solutions with $z=2$ were found, but, because the solutions
have the metric component $g_{++}$ positive in some regions of the
$SE_5$, the solutions were argued to be unstable.
In \cite{Donos:2009en}, supersymmetric solutions with $W=h=0$, $C\ne 0$ were constructed with
\be\label{betaeq}
\triangle_{SE}\beta=  \lambda_1\left(\lambda_{1}+2\right)\beta,\qquad d^\dagger\beta=0
\ee
where $\triangle_{SE}=dd^\dagger+d^\dagger d$ is the Hodge-deRahm operator on $SE_5$,
and give rise to solutions with $z=2+\lambda_1\ge 4$, with the bound achievable for any $SE_5$ space.
More specifically, the bound is achieved when $\beta$ is a one-form dual to a Killing vector on the $SE_5$
space; the class of such $z=4$ solutions using the one-form dual to the Reeb vector on the $SE_5$ space
were first constructed in \cite{Maldacena:2008wh}. It was also shown in \cite{Donos:2009en} that
one can combine these classes of solutions with
$h,C\ne 0$ (still with $W=0$), and providing that one can solve for $q$, $\beta$ so that $2+\lambda_1=1+\lambda_2/2$ then the
solutions have dynamical exponent $z=2+\lambda_1\ge 4$.

We now consider $W\ne 0$. This implies that $h\ne 0$ and we need to set
$\lambda_2=2(\lambda_3-1)$. In addition to
\reef{betaeq} we also need to solve
\begin{align}\label{z2eqns}
\triangle_{SE}\sigma=  \lambda_3\left(\lambda_{3}+2\right)\sigma,\qquad d^\dagger\sigma=0\nn
\nabla_{SE}^{2}q+4(\lambda_3^2-1)q=-\lambda^2_3|\sigma|^2_{SE}-|d\sigma|_{SE}^2~.
\end{align}
The solutions for which $\lambda_3=2+\lambda_1$ are invariant under Sch$\left( z\right)$with
$z=\lambda_3$. If $C\ne 0$ then since $\lambda_1\ge 2$, necessarily we have $z\ge 4$.

If we set $C=0$, which is needed to obtain supersymmetric solutions with $z=2$ for example, then we just need to solve
\reef{z2eqns}. The first equation implies that $z=\lambda_3\ge 2$, with the bound being saturated when $\sigma$
is a one-form dual to a Killing vector on the $SE_5$ space. A simple solution is obtained by taking $\sigma=c\eta$ for some constant $c$, where
$\eta$ is the canonical one-form dual to the Reeb vector on $SE_5$ and $q=-|c|^2$. This solution has $z=\lambda_3=2$ and
was first constructed in \cite{Herzog:2008wg,Maldacena:2008wh,Adams:2008wt}. Observe that for this solution $W=2cJ_{CY}$. Thus while $W$ is
$(1,1)$ it is not primitive and so this solution does not preserve any supersymmetry as previously pointed out in \cite{Maldacena:2008wh}.
On the other hand it is straightforward to construct solutions with $z=2$ that are supersymmetric. For example, we can take any Killing vector on
the $SE_5$ space that leaves invariant the Killing spinors on $SE_5$.
It is straightforward to construct such solutions explicitly when the metric for the $SE_5$ is known explicitly as it is for
the $S^5$, $T^{1,1}$ \cite{Candelas:1989js}, $Y^{p,q}$ \cite{Gauntlett:2004yd} and $L^{a,b,c}$ \cite{Cvetic:2005vk} spaces.
For the case of $S^5$ it is also easy to construct explicit solutions for all values of $z$
using spherical harmonics.
It is worth noting that the $z=2$ solutions for the $S^5$ case can have $q$ constant and negative
and hence do not suffer from the instability discussed in \cite{Hartnoll:2008rs}. This is easy to see since $W$ must be a constant
linear combination of the 15 harmonic two-forms on $\bbR^6$, $dx^i\wedge dx^j$, or, if we demand supersymmetry,
of the eight primitive $(1,1)$ forms and three $(2,0)$ forms. Then, in general, $q$ will be the sum of a negative constant
with a scalar harmonic on $S^5$ with eigenvalue 12.
It would be interesting to investigate the issue of
stability further for all of the new solutions we have constructed.
Some additional comments about the solutions are presented in appendix A.

\section{Solutions of $D=11$ supergravity}

We consider the ansatz for the bosonic fields of $D=11$ supergravity given by
\bea
ds^{2}&= & \Phi^{-{2}/{3}}\left[2dx^{+}dx^{-}+h\,\left(dx^{+}\right)^{2}+2dx^{+}C+dx_{1}^{2}\right]+\Phi^{{1}/{3}}ds^{2}\left(CY_{4}\right)\nn
G&= & dx^{+}\wedge dx^{-}\wedge dx_{1}\wedge d{\Phi}^{-1}+dx^{+}\wedge V+dx^{+}\wedge dx_{1}\wedge d\left(\Phi^{-1}{C}\right)
\eea
where $\Phi$, $h$ are functions, $C$ is a one-form and $V$ is a three-form all defined on\footnote{It is straightforward to also consider other
eight-dimensional special holonomy manifolds, but for simplicity we shall restrict our attention to $CY_4$.} the
Calabi-Yau four-fold, $CY_{4}$. Our conventions for $D=11$ supergravity
\cite{Cremmer:1978km} are as in \cite{gp}.
One finds that all the equations of motion are satisfied provided that
\bea
\label{begeqs}
\nabla^{2}_{CY}\Phi&= & 0\nn
d\ast_{CY}dC&= & 0\nn
dV=d\ast_{CY}V&= & 0\nn
\nabla^{2}_{CY}h&= & -|V|^2_{CY}
\eea
where $|V|^2_{CY}\equiv (1/3!)V^{ijk}V_{ijk}$ with indices raised
with respect to the $CY$ metric.
When $C=h=V=0$ we have the standard M2-brane class
of solutions with a transverse $CY_4$ space.

If we choose the three-form $V$ to only have $(2,1)$ plus $(1,2)$ pieces and be primitive on the $CY_4$
then the solutions generically preserve 
2 supersymmetries\footnote{As an aside, we note that we can also
add a closed, primitive $(2,2)$-form $F$ on $CY_4$ to the four-form flux while still preserving the same amount of
supersymmetry. This changes two of the equations to $\nabla^{2}_{CY}\Phi=-(1/2)|F|^2_{CY}$ and $d\ast_{CY}dC=V\wedge F$.},
which is enhanced to 
4 supersymmetries if the $CY_4$ us flat.
More specifically, we introduce the orthonormal frame
$e^+=\Phi^{-1/6}dx^+, e^-=\Phi^{-1/6}(dx^-+C+\frac{h}{2}dx^+), e^2=\Phi^{-1/6}dx^1,$ etc. and
choose positive orientation to be given by $e^{+-2}\wedge\rm{Vol}_{CY}$, where $\rm{Vol}_{CY}$ is the volume element on $CY_{4}$.
Consider first the special case that $C=h=V=0$. Then, as usual,
a non-flat $CY_4$ breaks 1/8 of the supersymmetry, and the harmonic function $\Phi$ can be added ``for free''
(the projection on the Killing spinors arising from the $CY_4$ automatically imply the projection $\Gamma^{+-2}\epsilon=-\epsilon$).
Switching on $C,h,V$ we find that generically we need to also impose $\Gamma^+\epsilon=0$ and $\Gamma^{ijk}V_{ijk}\epsilon=0$.
Note as an aside that we can ``skew-whiff'' by changing the sign of the four-form flux and
obtain solutions that generically don't preserve any supersymmetry (apart from the special case when $SE_7=S^7$).

We now specialise to the case that the $CY_{4}$ is a metric cone over a seven-dimensional Sasaki-Einstein manifold $SE_7$,
$ds^2(CY_4)=dr^2+r^2ds^2(SE_7)$. In order to get solutions with Sch$(z)$ symmetry we now set
\bea
\Phi&=&r^{-6}\nn
C&=& r^{\lambda_{1}}\beta\nn
h&=& r^{\lambda_{2}}q\nn
V&=& d\left(r^{\lambda_{3}}\tau\right)
\eea
where $q$ is a function, $\beta$ and $\tau$ are, respectively, a one-form and a two-form
on $SE_7$, and $\lambda_i$ are constants which we will take to be positive.
The full solution now reads
\bea
ds^{2}&= & r^{4}\left[2dx^{+}dx^{-}+r^{\lambda_{2}}q\,\left(dx^{+}\right)^{2}+2r^{\lambda_{1}}
\,dx^{+}\beta+dx_{1}^{2}\right]+\frac{dr^2}{r^2}+ds^{2}\left(SE_{7} \right)\nn
G&= & 6r^{5}\,dx^{+}\wedge dx^{-}\wedge dx_{1}\wedge dr+dx^{+}\wedge d\left(r^{\lambda_{3}}\tau \right)+
dx^{+}\wedge dx_{1}\wedge d\left(r^{6+\lambda_{1}}\beta\right)\nn
\eea
Generically, when $C,h,V\ne 0$, solutions with $2+\lambda_1/2=1+\lambda_2/4=\lambda_3/2\equiv z$ will be
Sch$(z)$ invariant. In particular, the scaling now acts as $(x^+,x^-,x_1,r)\to (\mu^zx^+,\mu^{2-z}x^-,\mu x_1,\mu^{-1/2}r)$.
Note that if we set $C=h=V=0$ then we have the standard $AdS_4\times SE_7$
solution. Generically, when $C,h,V\ne 0$, we still need to impose the projections mentioned above in
order to preserve supersymmetry. Thus, generically, half of the Poincar\'e supersymmetries
of the $AdS_4\times SE_7$ solution are preserved, while none of the special conformal supersymmetries are.
It would be interesting to explore 
special subclasses of solutions with enhanced supersymmetry, which occur,
for example, when 
the $CY_4$ is flat.

In \cite{Donos:2009en}, supersymmetric solutions with $C=V=0$, $h\ne 0$ were constructed with
\begin{align}
\nabla_{SE}^{2}q+\lambda_2(6+\lambda_2)q=0
\end{align}
and give rise to solutions with $z=1+\lambda_2/4\ge 5/4$, with the bound only achievable for
$SE_7=S^7$. In particular supersymmetric solutions with $z=2$ were found, but they
suffer from a similar instability to that found for the analogous type IIB solutions in \cite{Hartnoll:2008rs}.
In \cite{Donos:2009en}, supersymmetric solutions with $h=V=0$, $C\ne 0$ were constructed with
\be
\triangle_{SE}\beta=  \lambda_1\left(\lambda_{1}+4\right)\beta,\qquad d^\dagger\beta=0
\ee
and give rise to solutions with $z=2+\lambda_1/2\ge 3$, with the bound achievable for any $SE_7$ space.
More specifically, the bound is achieved when $\beta$ is a one-form dual to a Killing vector on the $SE_5$
space; and one can always choose the one-form dual to the Reeb vector on the $SE_7$ space.
It was also shown in \cite{Donos:2009en} that one can combine these classes of solutions with $C,h \ne 0$,
(still with $V=0$), and providing that one can choose $4+2\lambda_1=\lambda_2$ then they have dynamical
exponent again with $z=2+\lambda_1/2\ge 3$.

We now consider $V\ne 0$. This implies $h\ne 0$ and we need to set
$\lambda_2=2(\lambda_3-2)$. In addition to
\reef{betaeq} we also need to solve
\begin{align}\label{z3eqns}
\triangle_{SE}\tau=  \lambda_3\left(\lambda_{3}+2\right)\tau,\qquad d^\dagger\tau=0\nn
\nabla_{SE}^{2}q+4(\lambda_3-2)(\lambda_3+1)q=-\lambda^2_3|\tau|^2_{SE}-|d\tau|_{SE}^2~.
\end{align}
The solutions for which $\lambda_3=4+\lambda_1$ are invariant under Sch$\left( z\right)$ with
$z=2+\lambda_1/2=\lambda_3/2$. If $C\ne 0$ then necessarily we have $\lambda_1\ge 2$ and hence $z\ge 3$.

If we set $C=0$ then we just need to solve \reef{z3eqns}. Let us illustrate with some simple solutions
when $SE_7=S^7$. In fact it is easiest to directly solve \reef{begeqs}.
For example, if we let $z^a$ be standard complex coordinates on $\bbR^8$, with K\"ahler form
$\omega=(i/2)dz^a\wedge d\bar z^a$ we can take
$V=c\,dz^1d\bar z^2 d\bar z^3 +c.c.$, where $c$ is constant,
which obviously has only $(1,2)$ and $(2,1)$ pieces and
is primitive, and $h=-c^2\,r^2$ (setting a possible solution of the homogeneous equation in \reef{begeqs} to zero).
This gives a supersymmetric solution with $\lambda_3=3$ and hence $z=3/2$. In particular we note that the metric component $g_{++}$ is always negative.
A simple solution with $z=2$ is obtained by splitting $\bbR^8=\bbR^4\times \bbR^4$ and considering a sum of terms which are $(1,1)$ and primitive
on one factor with a factor $dx^i$ on the other: 
\bea
V&=&c\Big\{ \left[x^{1}\, (dx^{1}\wedge dx^{2}-dx^{3}\wedge dx^{4})+x^3(dx^{2}\wedge dx^{3}-dx^{1}\wedge dx^{4})\right]\wedge dx^5\nn
&+&\quad \left[x^{2}\, (dx^{1}\wedge dx^{2}-dx^{3}\wedge dx^{4})+x^4(dx^{2}\wedge dx^{3}-dx^{1}\wedge dx^{4})\right]\wedge dx^6\nn
&+& \quad\left[x^{5}\, (dx^{5}\wedge dx^{6}-dx^{7}\wedge dx^{8})+x^7(dx^{6}\wedge dx^{7}-dx^{5}\wedge dx^{8})\right]\wedge dx^1\nn
&+&\quad \left[x^{6}\, (dx^{5}\wedge dx^{6}-dx^{7}\wedge dx^{8})+x^8(dx^{6}\wedge dx^{7}-dx^{5}\wedge dx^{8})\right]\wedge dx^2\Big\}~.\nn
\eea
Solving for $h$ (and setting to zero a solution of the homogeneous equation in \reef{begeqs}) we get
\begin{equation*}
h=-\frac{c^2}{20}\,r^{4}~.
\end{equation*}
For this solution, the metric component $g_{++}$ is again always negative.
Clearly there are many additional simple constructions for the $S^7$ case
that could be explored as well as for the more general class of other explicit $SE_7$ metrics.

\subsection*{Acknowledgements}
JPG is supported by an EPSRC Senior Fellowship and a Royal Society Wolfson Award.

\appendix

\section{Comments on solving \reef{z2eqns}}
Here we make a few further comments concerning solving \reef{z2eqns} (which also have obvious analogues for
solving \reef{z3eqns}).
To solve \reef{z2eqns}, we first solve the first line for $\sigma$ and then substitute into the second.
It is illuminating to expand out the source term in the right hand side of the equation in the second line
using a complete set of scalar harmonics on the $SE_5$ space:
\be
-\lambda^2_3|\sigma|^2_{SE}-|d\sigma|_{SE}^2 = \sum_{I_l}a_{I_l} Y^{I_l}
\ee
where $\nabla^2_{SE}Y^{I_l}=-l(l+4)$, corresponding to the harmonic function $P^{I_l}=r^lY^{I_l}$ on the $CY_3$ cone.
We then find
\begin{equation}\label{qsol}
q=\sum_{I_l}\frac{a_{I_l}}{4\lambda_3^{2}-(l+2)^2}Y^{I_l}+q_{0}~.\end{equation}
In this expression we have allowed for the possibility of an arbitrary
solution to the homogeneous equation, $q_0$, assuming it exists. The point is
that the relevant putative eigenvalue for $q_0$ is fixed by the eigenspectrum of the Laplacian acting on one-forms.
For the special case when $SE_5=S^5$, for example, there is always such a possibility of adding a solution to the homogeneous equation.
Another point to notice about \reef{qsol} is that it only makes sense providing that the coefficient
$a_{I_l}=0$ whenever $2\lambda_3=l+2$.

For the special case when $SE_5=S^5$, not only is this coefficient zero but the sum appearing in \reef{qsol}
is a finite sum terminating at $l=2\lambda_3-4$. To see this we observe that
\begin{equation}
a_{I_l}\propto\int_{S^{5}}Y^{I_l}(\lambda^2_3|\sigma|^2_{SE}+|d\sigma|_{SE}^2)\end{equation}
which can be recast as an integral on the flat cone $\bbR^6$
\begin{equation}
a_{I_l}\propto\int_{\bbR^{6}}e^{-r^2}P^{I_l}\, W^{ij}W_{ij}\label{eq:int_cone}\end{equation}
where for $S^5$, $r^2=\sum_i x^i x^i$ and
\be
P^{I_l}= C_{i_{1}\ldots i_{l}}^{I}x^{i_{1}}\cdots x^{i_{l}}
\ee
with $C_{i_{1}\ldots i_{l}}^{I}$ defining the scalar harmonics on $S^5$.
To proceed we write $W$ as
\be
W=  C_{j;ki_{1}\ldots i_{\lambda_3-1}}^{J}x^{i_{1}}\cdots x^{i_{\lambda_3-2}}\, dx^{j}\wedge dx^{k}
\ee
where $C_{j;i_{1}\ldots i_{\lambda_3-1}}^{J}$ define the vector spherical harmonics on $S^5$.
In carrying out the integral \reef{eq:int_cone} we will get all possible contractions of the
$l$ indices of the scalar spherical harmonic  $C_{i_{1}\ldots i_{l}}^{I}$ with some of the $2\lambda_3-4$ indices
\begin{equation}
C_{\left[j;k\right]i_{1}\ldots i_{\lambda_3-2}}^{J}C_{\left[j;k\right]i'_{1}\ldots i'_{\lambda_3-2}}^{J}\end{equation}
In particular, since the tensor defining the scalar harmonic is traceless,
we conclude that 
the $a_{I_l}$ are zero for all $I_l$ with $l> 2\lambda_3-4$. 

Let us now consider this issue for a general $SE_5$ space, but in the special case when $\sigma$ is a one-form
dual to a Killing vector on $SE_5$ corresponding to $\lambda_3=2$ and hence $z=2$.
As above, we have \reef{eq:int_cone}.
Write
\be
W=d(r^{2}\sigma)\equiv dT
\ee
and observe that on the $CY_3$ cone $\nabla_i T_j=\nabla_{[i}T_{j]}$ and that
$\nabla^2_{CY}T_i=0$. We then compute
\bea\label{lastone}
a_{I_l}&\propto&\int_{CY}e^{-r^{2}}P^{I_l}\, W^{ij}W_{ij}\nn
&= &4\int_{CY}e^{-r^{2}}P^{I_l}\left(\nabla^{i}T^j\right)\left(\nabla_{i}T_{j}\right)\nn
& =&2\int_{CY}\nabla^2_{CY}\left(e^{-r^{2}}P^{I_l}\right)T^{2}\nn
 & =&2\int_{CY}e^{-r^{2}}\left(-4r\partial_{r}P^{I_l}-12P^{I_l}+4r^{2}P^{I_l}\right)T^{2}\nn
 & =&4\int_{CY}e^{-r^{2}}\left(-l+2\right)P^{I_l}T^{2}~.
 \eea
In getting to the last line one needs to take into account the $r^5$ factor in the measure and use
\be
\int_{0}^{\infty}r^{n+2}e^{-r^{2}}dr=\frac{n+1}{2}\int_{0}^{\infty}r^{n}e^{-r^{2}}dr~.
\ee
We thus conclude from \reef{lastone} that the problematic coefficient $a_{I_l}$ in \reef{qsol} when $l=2\lambda_3-2=2$ again vanishes for
this class of solutions.

\end{document}